\documentclass[10pt,letterpaper]{article}
\usepackage[top=0.85in,left=2.75in,footskip=0.75in]{geometry}

\usepackage{comment}
\usepackage{amsmath,amssymb}

\usepackage{changepage}

\usepackage[utf8x]{inputenc}

\usepackage{textcomp,marvosym}

\usepackage{cite}

\usepackage{nameref,hyperref}

\usepackage[right]{lineno}

\usepackage{microtype}
\DisableLigatures[f]{encoding = *, family = * }

\usepackage[table]{xcolor}

\usepackage{array}
\usepackage{mathtools}
\newcolumntype{+}{!{\vrule width 2pt}}

\newlength\savedwidth

\newcommand\thickhline{\noalign{\global\savedwidth\arrayrulewidth\global\arrayrulewidth 2pt}%
\hline
\noalign{\global\arrayrulewidth\savedwidth}}

\raggedright
\usepackage[aboveskip=1pt,labelfont=bf,labelsep=period,justification=raggedright,singlelinecheck=off]{caption}

\makeatletter
\renewcommand{\@biblabel}[1]{\quad#1.}
\makeatother

\usepackage{lastpage,fancyhdr,graphicx}
\usepackage{epstopdf}
\fancyheadoffset[L]{2.25in}
\fancyfootoffset[L]{2.25in}
\begin{document}
\vspace*{0.2in}

\begin{flushleft}
{\Large
\textbf\newline{Novel Non-Invasive In-house Fabricated Wearable System with a Hybrid Algorithm for Fetal Movement Recognition } 
}
\newline
\\
Upekha Delay\textsuperscript{1,2\Yinyang},
Thoshara Nawarathne\textsuperscript{1\Yinyang},
Sajan Dissanayake\textsuperscript{1,3\textcurrency},
Samitha Gunarathne\textsuperscript{1},
Thanushi Withanage\textsuperscript{1},
Roshan Godaliyadda\textsuperscript{1},
Chathura Rathnayake\textsuperscript{2},
Parakrama Ekanayake\textsuperscript{1},
Janaka Wijayakulasooriya\textsuperscript{1},
\\
\bigskip
\textbf{1} Department of Electrical and Electronic Engineering, Faculty of Engineering, University of Peradeniya, Peradeniya [20400], Sri Lanka.
\\
\textbf{2} Department of Obstetrics and
Gynacology, Faculty of Medicine, University of Peradeniya, Peradeniya [20400], Sri Lanka.
\\
\bigskip

%
%
\Yinyang These authors contributed equally to this work.





* Corresponding author. upekha.delay@eng.pdn.ac.lk

\end{flushleft}
\section*{Abstract}

Fetal movement count monitoring is one of the most commonly used methods of assessing fetal well-being. While few methods are available to monitor fetal movements, they consist of several adverse qualities such as unreliability as well as the inability to be conducted in a non-clinical setting. Therefore, this research was conducted to design a complete system that will enable pregnant mothers to monitor fetal movement at home. This system consists of a non-invasive, non-transmitting sensor unit that can be fabricated at a low cost. An accelerometer was utilized as the primary sensor and a micro-controller based circuit was implemented. Clinical testing was conducted utilizing this sensor unit. Two phases of clinical testing procedures were done and readings from more than 120 pregnant mothers were taking. Validation was done by conducting an abdominal ultrasound scan which was utilized as the ground truth during the second phase of the clinical testing procedure. A clinical survey was also conducted in parallel with clinical testings in order to improve the sensor unit as well as to improve the final system. Four different signal processing algorithms were implemented on the data set and the performance of each was compared with each other. Consequently, the most feasible as well as the best performing algorithm was determined and it was utilized in the final system. Furthermore, a mobile application was also developed to be used with the sensor unit by pregnant mothers. Finally, a complete end to end method to monitor fetal movement in a non-clinical setting was presented by the proposed system.


\section*{Introduction}
During pregnancy, the main aim of the parents, as well as the obstetricians, is to maintain the health of the fetus as well as the mother. Obstetricians use different methods to assess fetal health. Among them monitoring fetal movement is the most commonly used method. It is also the simplest and the most economical method available\cite{1}.

Several studies have been carried out to classify the different types of fetal movements \cite{2}. It was mentioned that fetal movements can be classified based on the amplitude and the speed of the specific activity. The amplitude could be weak vs strong and the length of the activity could be short vs sustained. In a study conducted \cite{3} seven different fetal movement types were identified using Ultrasound imaging and Doppler Ultrasound. They were Startles, General movements, Hiccups, Fetal breathing movements, Isolated arm or leg movement, Twitches and Clonic movements. It was also noted that there is a striking similarity between the observed movements and the movements observed in a baby after birth. 

In a survey conducted 99.9\% of pregnant mothers reported that it was important to feel and count baby movements\cite{7}. Fetal movement patterns of each trimester of the pregnancy may differ from each other as well as from each fetus and mother. Studies have shown that the perception of decreased fetal movement is associated with stillbirth\cite{1.5}. Any difference in the fetal movement pattern can indicate that the fetus is unwell. The change in the pattern could be reduced fetal movement, weak fetal movement or intense fetal movement. Hence, change in fetal movement pattern may be a sign of an unhealthy fetus \cite{3}. It was also shown that by timely reporting to health care providers when experiencing a decreased fetal movement may prevent perinatal morbidity and mortality\cite{2}. A study carried out on 305 women who experienced reduced fetal movement after 28 weeks of gestation showed that 22.1\% pregnancies ended in complications such as small-for-gestational-age infants\cite{4}. Another study reported 54.7\% cases of stillbirth resulting women present who experienced reduced fetal movements\cite{5}. In another study 20 out of 23 growth-restricted fetuses were identified prior to birth by monitoring fetal movement counting where as only 12 out of 20 growth-restricted fetuses were identified with standard antenatal care without fetal movement counting \cite{6}. Also, several studies have been conducted to identify fetal movement patterns in each trimester as well as fetal movement patterns in of mothers\cite{4}\cite{5}\cite{6}. Another study conducted by the Royal College of Obstetricians and Gynecologists has highlighted the lack of studies of fetal movement patterns\cite{8}. Therefore, it can be concluded that monitoring fetal movement patterns play a major role in fetal well-being. Furthermore, further studies need to be conducted on how fetal movement patterns effect the well-being of the fetus.

Currently, fetal movements can be quantified by conducting an ultrasound scan or an MRI scan\cite{9}\cite{10}. These can only be conducted in a clinical setup and can only be done by a trained technician \cite{9}. In a study, an automated method of analysing fetal growth using 2D ultrasound was developed \cite{10}. This was done due to the lack of trained sonographers in developing countries. This indicates that there's a lack of trained technicians to conduct these tests. As a result, these tests are expensive and can only be conducted in a short window of time.

Few research studies were conducted on monitoring fetal movement via non-invasive techniques. A study conducted using a single accelerometer placed on the mother's abdomen employed a threshold signal processing method \cite{11}. This study concluded that the thresholding method employed to identify fetal movement performed poorly. Also, it was found that the acceleration signals are corrupted by maternal movements such as laugh, cough and hiccup. Therefore, a more complex signal analysis method needs to be employed. Another study \cite{12} employed a capacitive accelerometer to monitor fetal movements while the mother was asleep. In the initial part of the experiment, an ultrasonographer was used parallel to the device to validate data acquired. During this experiment, three types of fetal movements were recorded and their positive hit rates were calculated. Gross movements had a positive hit rate of 38.5\% - 23.5\% depending on the fetal age. Similarly, isolated limb movements had a positive hit rate of 5\% to 13\% and breathing movements had a lower positive hit rate of 4.3\% - 22.8\%. Another study \cite{13} was conducted using acoustic sensors and accelerometer sensors.This was also conducted in a clinical setup with ultrasound validation. They were able to discriminate fetal startle movements from general movements with 72.1\% accuracy. Several other research studies also have been conducted on this topic\cite{int1,int2,int3}.

In this study, we investigate whether an accelerometer sensor can be used to monitor the fetal movement count in a non clinical setting.

\section*{Materials and methods}
\subsection*{Hardware}
Few studies were conducted on wearable sensor-based fetal monitoring devices \cite{11}\cite{12}\cite{14}.  The most common sensors used in previous studies were accelerometers and acoustic sensors. Furthermore, it was mentioned in a study conducted that it is possible to monitor fetal movements as well as heartbeat using an accelerometric sensor. However, it was also mentioned that this is realizable only after the 30$^{th}$ week of gestation. This is due to the lack of strength of the signals generated when the fetus is in the early gestational stages\cite{16}. Therefore, it was decided to that an accelerometric sensor should be used to capture data.

A device was initially developed to acquire signals from pregnant mothers. When selecting a sensor for this device several factors were considered: One of the main considerations made was using a non-invasive sensor to acquire data. This was to reduce the impact of the device on the fetus as well as on the mother. Furthermore, the main objective of this was to come up with a wearable device which can be used by mothers daily. Therefore the ergonomics of the sensor also played a major role. The sensor to be selected should be light in weight and small in size. Also, it should be easy to wear on the mother's abdomen. Since this device should be commercially sold the cost also play a major role.  

Considering all these factors our research team decided to use the sensor MPU 9250. 
It is a multi-chip module which houses a 3-Axis accelerometer and a 3-Axis gyroscope. It has inbuilt analogue to digital converters to convert the signals received from the accelerometer and the gyroscope. The data received from the sensor is transferred to a removable micro SD card via a microcontroller. This enables the device to operate independently of a computer and also eliminates the need for a wireless transfer method which may have adversary effects on the fetus\cite{14}.

A previous study was conducted using accelerometers to decide the optimum number of sensors to be used and the sensor positioning. During this study, five accelerometers were used and they were positioned on the abdomen with the navel as the centre mark. Additionally, a reference sensor was placed on the back. Readings of 6 pregnant mothers whose gestational age was from 30 upwards were taken. The maternal perception was considered to be the ground truth while only the fetal kicks were taken into consideration. In this test, the increase in the number of sensors did not have a considerable effect on the positive predictive value. While the positive predictive value increased when the number of sensors were increased, the difference was not notable\cite{15}. Therefore, it was decided that a single sensor should be used in the setup. This idea was further supported by the notion that this will reduce the computational capacity required as well as the size of the data recorded which in turn will improve the speed of analysis as well as data transfer.

The clinical testing procedure was conducted in two phases. During the initial phase mothers perception of fetal movement was considered to be the ground truth. However, during the second phase, ultrasound readings were utilized for validation and as the ground truth. During both of these phases, a belt-like device was used to collect data. This can be seen in Fig 1.
\clearpage

\begin{figure}[!h]
\includegraphics[width=\textwidth]{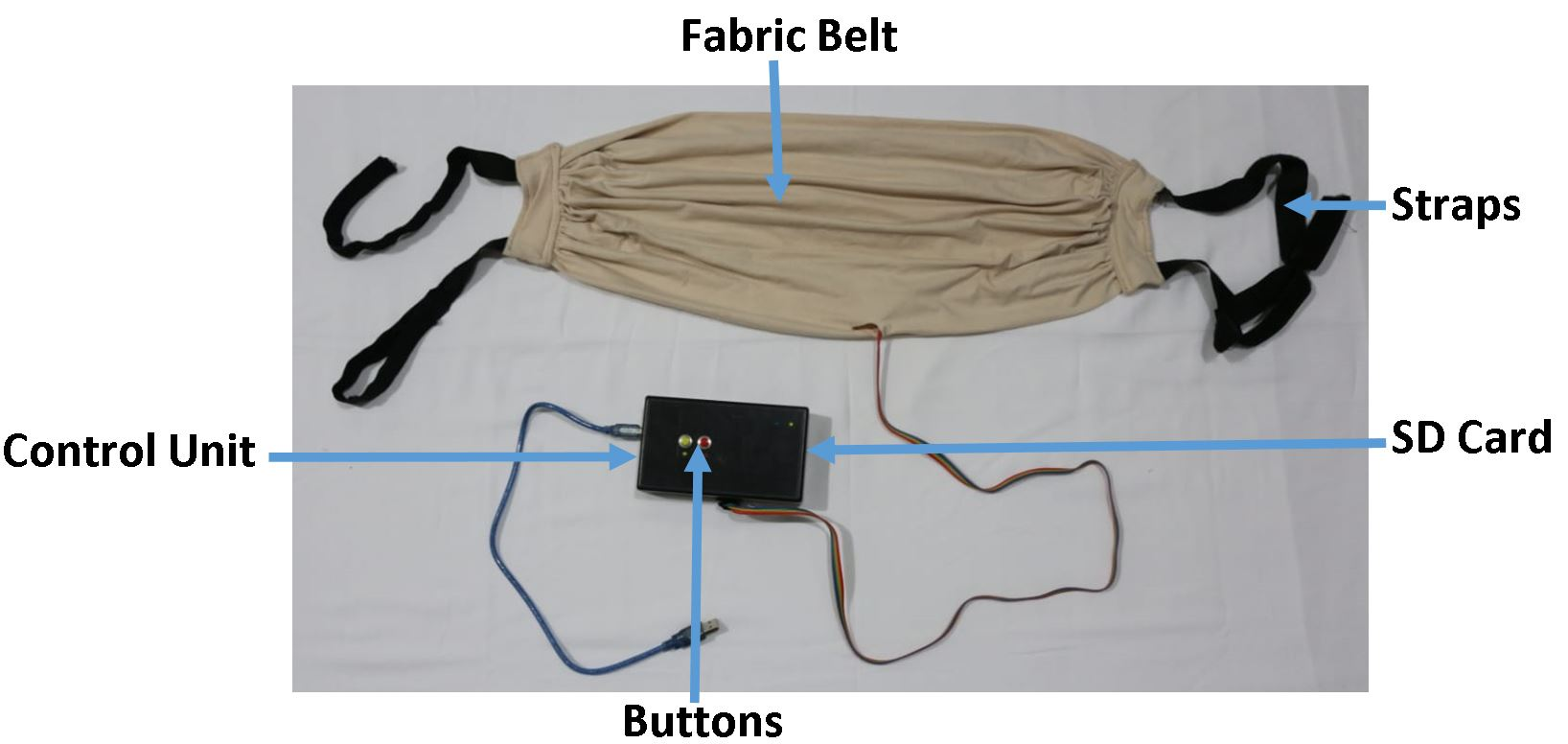}
\caption{{\bf The in-house designed and fabricated device.}
}
\label{fig:device}
\end{figure}

Initially the sensor, MPU 9250 was embedded into a rubber sole. This was to adhere the sensor on to the abdomen securely as well as to prevent any discomfort to the mother due to the sharp edges of the sensor. Then this rubber sole was stitched into a fabric belt. The choice of the material to the fabric belt was also done considering the mother's comfort. Several factors such as the materials ability to absorb perspiration, flexibility, how well it moulds to the mother's abdomen as well as the colour of the material were considered when choosing the material. This selection was made by considering the responses of several pregnant mothers who were interviewed during the design process. The dimensions of this belt were designed in such a way so that it could be worn for an extended time period. Furthermore, the sensor was placed at the centre of the belt in order to obtain a uniform sensor positioning on every mother.

During the clinical testing period, the microcontroller as well as the SD card were housed in a separate box. Two buttons were also included in this setup. This housing can be seen in Fig 1.
One of the buttons was used to obtain mothers perception of fetal movements. When taking readings the mother was advised to press the button on the device whenever a fetal movement is felt. The other button was used to record maternal movements such as laugh, cough and hiccups.

\subsubsection*{Ethical clearance}
This study was approved by the Ethics Review Committee, Faculty of Medicine, University  of Peradeniya.Approval was granted to conduct  research project No. 2018/EC/43 entitled "Fetal movement analysis for condition monitoring" at the Teaching Hospital, Peradeniya.

\subsection*{Clinical Tests}

The clinical testing procedure was conducted in two phases. During phase one mother's perception of fetal movements were considered to be the ground truth. During this phase, the design and development of the device were also done. Some adjustments were made to the device during this period according to the feedback received from pregnant mothers. At the conclusion of phase one, a proper device was designed and a proper method of taking readings was developed. Then at phase two ultrasound readings were considered to be the ground truth. The device which was finalized in phase one was utilized in this phase to obtain readings.

\textbf{Phase 1:} During this phase readings from 127 women impatient at Peradeniya Teaching hospital, Sri Lanka were taken. The gestational age of the group varied from 28 weeks to 40+ weeks and most were singleton pregnancies. A twin pregnancy as well as a quadruplets pregnancy was also recorded. However, only the readings from singleton pregnancies were initially considered. Furthermore, the occurrence of fetal movements and the occurrence of maternal movements were recorded utilizing the button system mentioned above.

\textbf{Phase 2:} During this phase readings from 15 mothers were taken and all of them were impatient at Peradeniya Teaching Hospital, Sri Lanka. Similar to phase one the gestation period of these mothers varied from 28 weeks to 40+ weeks and all the pregnancies were singleton. Furthermore, during this phase, each mother underwent an abdominal ultrasound, and the fetal movements were monitored and recorded by a trained technician while the developed device recorded accelerometric data. Other maternal movements were also recorded. Moreover, the mother's perception was also recorded.

During both these phases pregnant mother's written consent was obtained prior to acquiring data and a thorough explanation on the process of acquiring data as well as the final target of the research was given. Then basic details about the mother as well as the fetus were recorded. The data collected were, mother's age, fetal age, fetal gender, number of previous pregnancies, expected delivery method and additional comments. After that the belt containing the sensor was placed around the mother's abdomen and secured to reduce the movements of the belt relative to the mother's abdomen. Then the mother was advised to stay in a comfortable position. In phase one most mothers chose to lay down while a few were sitting up. However, during phase two every mother had to lay down in order to accommodate the ultrasound probe as well as the device. Each session was approximately 20 minutes long and for every mother, a single session was conducted. Therefore, more than 5 hours of readings were acquired. 

\subsubsection*{Class Identification}
When considering the maternal movements such as laugh, cough and hiccup, it was observed that maternal laugh is the most frequent maternal movement. Furthermore, it has the most similarities to the fetal movement signal observed. Therefore, during phase one three classes were introduced. They were fetal movement, maternal laugh and mothers respiratory movements. During phase two, three types of fetal movements were observed. They are limb movements, rotations and whole body movements. From these three types, only the fetal limb movements were considered as fetal movements when conducting the analysis. Therefore, the three classes identified during phase two are fetal limb movements, maternal laugh and mother's respiratory movements. The frequency of each class occurrence can be observed in Table 1.

\begin{table}[!ht]
\centering
\caption{
{\bf Number of occurrences of the three classes with gestation age.}}
\begin{tabular}{|l+l+l+l|l|l|l|}
\hline

$ \bf Fetal Age (Weeks) $ & \bf Class 1 & \bf Class 2 & \bf Class 3 \\ \thickhline
$ \bf 27 - 31 $ & 174 & 35 &  263\\ \hline
$ \bf 32 - 35 $ & 265 & 78  & 360\\ \hline
$ \bf 36 - 40+$ & 583 & 163 & 954\\ \thickhline
$ \bf Total$ & \bf 1022 & \bf 276 & \bf 1563 \\ \hline
\end{tabular}
\begin{flushleft} The total number of occurrences of each class can be observed. The fetal movement realizations are classified as Class 1, Maternal laugh realizations are classified as Class 2 and Maternal respiratory movement realizations are classified as Class 3.
\end{flushleft}
\label{realizations}
\end{table}

\subsubsection*{Clinical Survey}
The end goal of this research was to design and fabricate a fetal movement monitoring system which can be used by pregnant mothers at home. Therefore, a survey was conducted alongside clinical tests. The results obtained from this was utilized when designing the end system. Following observations were made during the survey. Every mother interviewed kept track of fetal movement during the pregnancy as advised by their obstetrician. However, this was done by keeping track of whether a fetal movement occurred during each hour of the day. The general opinion of the mothers was that this method was unreliable and a nuisance. Furthermore, they were advised to observe fetal moments immediately after consuming food. On average, a mother with a healthy fetus undergoes three ultrasound scans during the gestational period and undergoes a Cardiotocography (CTG) scan daily when impatient at the hospital. More than 80\% of the mothers interviewed had a favourable opinion about the proposed system of fetal movement monitoring, while approximately 15\% were indecisive. Only less than 5\% of mothers had an unfavourable opinion about the proposed system. Moreover, almost every mother interviewed reacted favourably to the idea of including a mobile application to the proposed system.

\subsection*{Observations}

The data obtained from the sensor were stored in the micro SD card which was later transferred and analysed. Prior to conducting analysis, several important observations were made. The accelerometer in the sensor measures the acceleration along the three axes: X-Axis, Y-Axis and Z-Axis. Z-Axis records the acceleration variation normal to the mother's abdomen while X and Y axes record the acceleration variation along the plane of the abdomen. Raw time-domain data obtained can be observed in Fig 2.

\begin{figure}[!h]
\includegraphics[width=\textwidth]{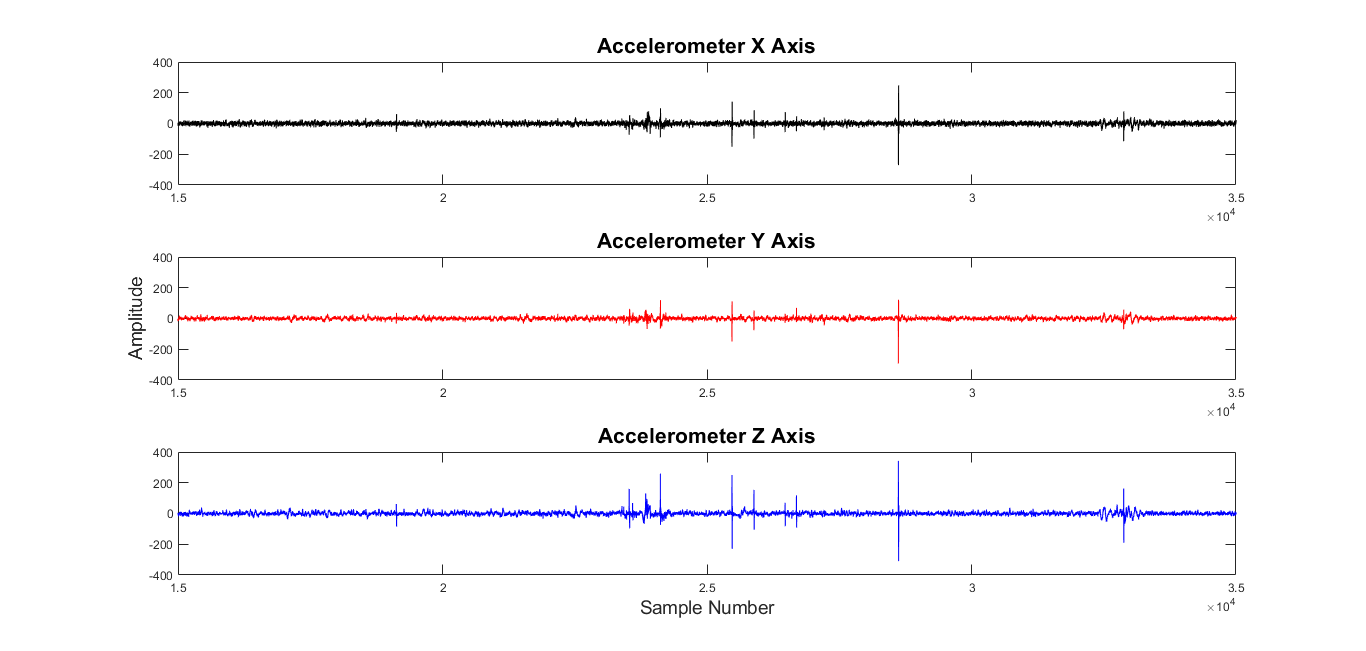}
\caption{{\bf The accelerometric data along the three axes, X-Axis, Y-Axis and Z-Axis.}\
}
\label{fig:xyz}
\end{figure}

From Fig 2 it can be observed that the variation along the Z-Axis is more prominent than the variation along the other two axes. This argument can be further solidified by observing Fig 3. In Fig 3 the variation of data point in the 3 dimensional space can be observed during a fetal movement realization. In it, it can be observed that the most prominent variation is along the Z-Axis. Therefore, only Z-Axis data were utilized when conducting the analysis. Furthermore, this results in a reduction of computational capacity requirement, which in turn, is favourable when conducting complex algorithms.

\begin{figure}[!h]
\includegraphics[width=\textwidth]{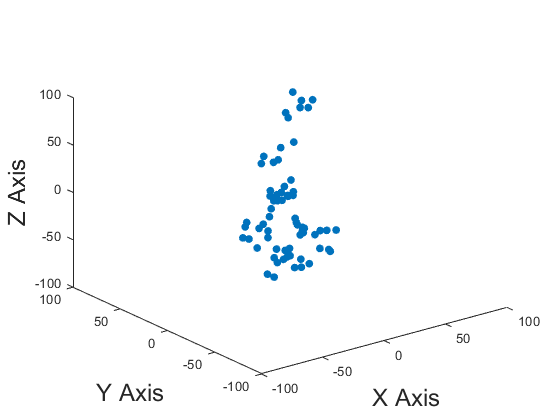}
\caption{{\bf The variation of data point in the 3 dimensional space during a fetal movement realization.}\
}
\label{fig:xyz-point}
\end{figure}

During both phases of clinical testing, the realizations were segmented into three classes. They are fetal movement, maternal laugh and maternal respiratory movements. Visually the time-domain representation of the fetal movement and maternal laugh have similarities while clear discrimination can be made between these two and maternal respiratory movements. This can be observed in Fig 4. 

\begin{figure}[!h]
\includegraphics[width=\textwidth]{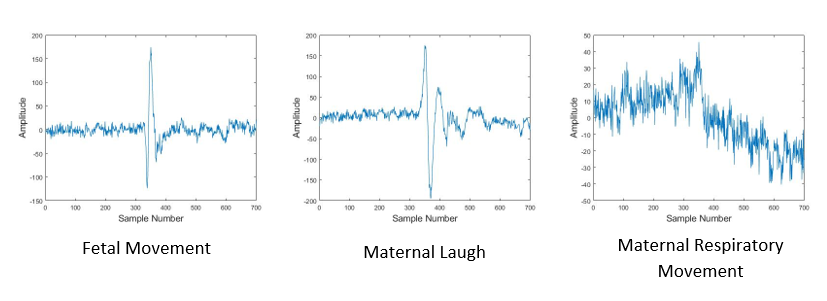}
\caption{{\bf The accelerometric data of realizations from the three classes.}\
}
\label{fig:three-realizations}
\end{figure}

During the second phase of the clinical testing procedure, an abdominal ultrasound was conducted to validate the results. The time-domain signal of a singleton pregnancy observed by the sensor and the fetal movement identified by each method can be observed in Fig 5. Furthermore, a similar diagram of the time domain signal of a breached fetal can be observed in Fig 6.

\clearpage

\begin{figure}[!h]
\includegraphics[width=\textwidth]{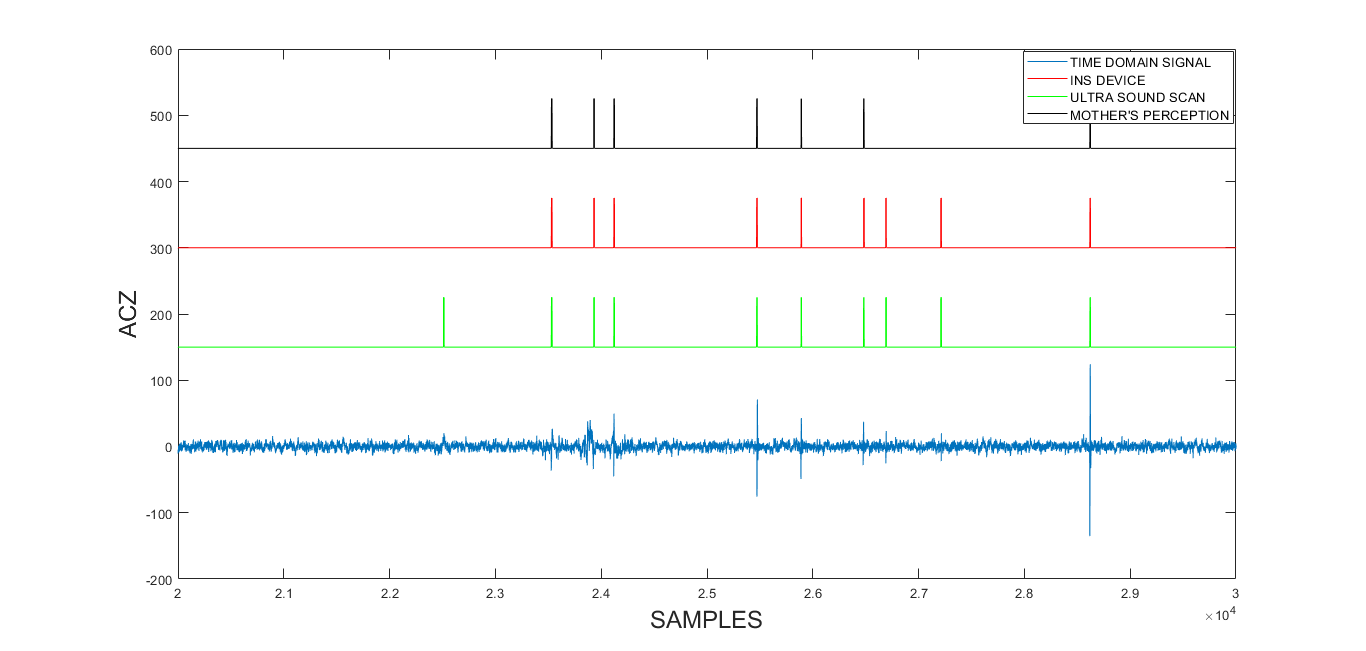}
\caption{{\bf The time-domain signal of a singleton pregnancy observed by the sensor and the fetal movement identified by each method.}\
}
\label{fig:methods-normal}
\end{figure}

\begin{figure}[!h]
\includegraphics[width=\textwidth]{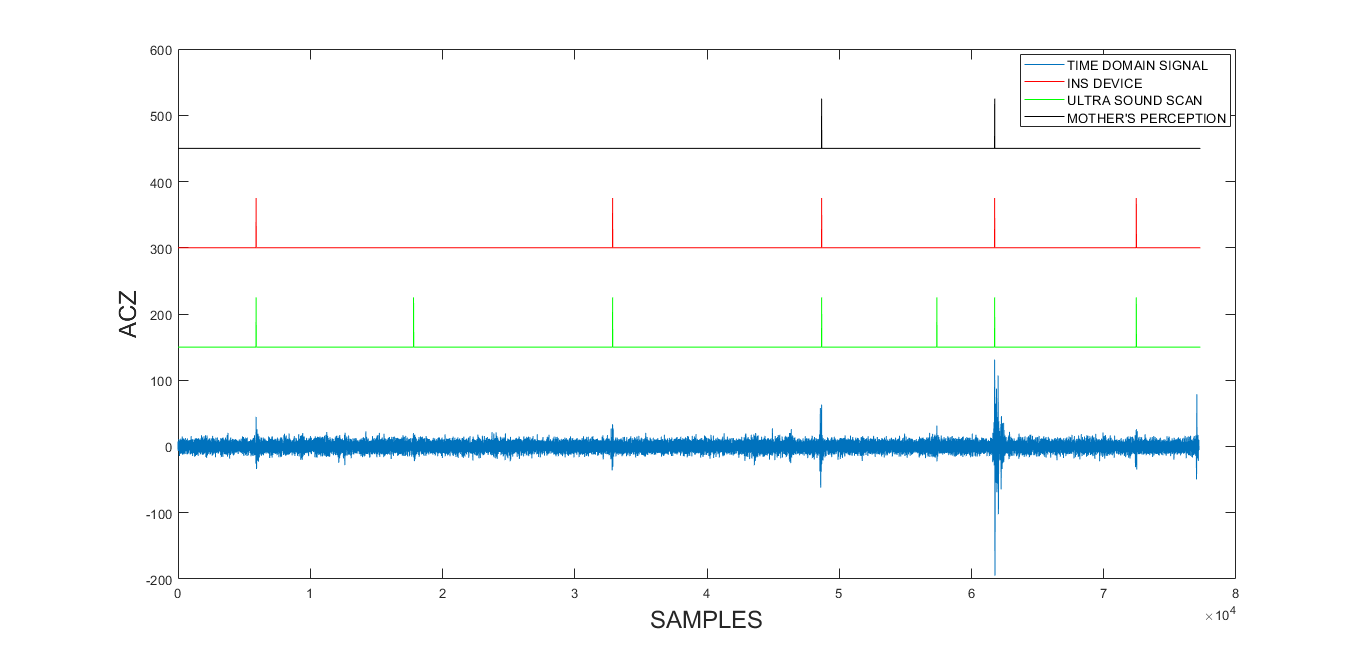}
\caption{{\bf The time-domain signal of a singleton breached pregnancy observed by the sensor and the fetal movement identified by each method.}\
}
\label{fig:methods-breached}
\end{figure}

If the abdominal ultrasound is to be considered as the ground truth, it can be observed that some fetal movements were not felt by the mother as well as the device. It can be observed from the Fig 5 that in the normal singleton pregnancy most fetal movements identified by the ultrasound were also observed by the mother and the device. However, when the fetus is breached, most of the fetal movements identified by the ultrasound were not observed by the mother but they were observed by the device. Therefore, it can be derived that this proposed system can be utilized to identify and monitor fetal movement which can not be felt by the mother. A similar analysis was conducted for the entire data set obtained during the second phase. It was observed that 65.56\% of realizations were identified by all three methods. However, 15.56\% of realizations were identified by the ultrasound and the device and 18.89\% of realizations were only identified by the ultrasound. The summery of this data can be seen in Table 2. Furthermore the mothers were not able to identify fetal movements which were not captured by the ultrasound scan and all the movements felt by the mothers were also captured by the device.

\begin{table}[!h]
\centering
\caption{
{\bf Number of realizations identified and observed by each method during the second phase of clinical  testing.}}

\begin{tabular}{|l+l|l|l|l|}
\hline
$ \bf Ultra $ & \bf Device & \bf Mother's  & \bf Number of & \bf Total \\
$ \bf Sound $ & \bf & \bf Response & \bf Detected Kicks & \bf Percentage(\%)\\ \thickhline
$ 1 $ & 0 & 0 & 17 & 18.89\\ \hline
$ 1 $ & 1 & 0 & 14 & 15.56 \\ \hline
$ 1 $ & 1 & 1 & 59 & 65.56 \\ \hline

\end{tabular}
\begin{flushleft} During the second phase of clinical testing three methods were utilized to identify the occurrence of a fetal movement. They are: the abdominal ultrasound, the in-house fabricated device and mother's response. In the table '1' indicates that the relevant method was able to identify the fetal movement and '0' indicates that the method was not able to identify the fetal movement. For instance the first data row of the table indicated the number of realizations captured only by the abdominal ultrasound. The second row indicate the number of fetal movement captured by the ultra sound and the in-house fabricated device. And the final row indicate the fetal movements captured by all three methods. 
\end{flushleft}
\label{methods}
\end{table}

\subsection*{Signal Analysis}

Several combinations of signal processing algorithms were administered in order to obtain an optimum algorithm. The main objective of the algorithm was to successfully compute the number of fetal movement occurrences in a single session and to minimize the probability of false positives as these could result in extensive adversarial effects. The entire algorithm was implemented using MATLAB and later implemented on Android studio in order to implement this algorithm on a smart phone.

\textbf{Pre processing :} Initially the signals obtained were observed and it was noticed that due to maternal movements the signal has shifted as seen in Fig 7. Furthermore maternal breathing has introduced a periodic noise component which can also be observed in Fig 8. Therefore, the raw time domain signal was initially sent through a high pass filter which was utilized in a similar research study done previously \cite{2}. As it can be observed in Fig 8 when the filter is applied maternal movements as well as maternal breathing noise were eliminated.

\begin{figure}[!h]
\includegraphics[width=\textwidth]{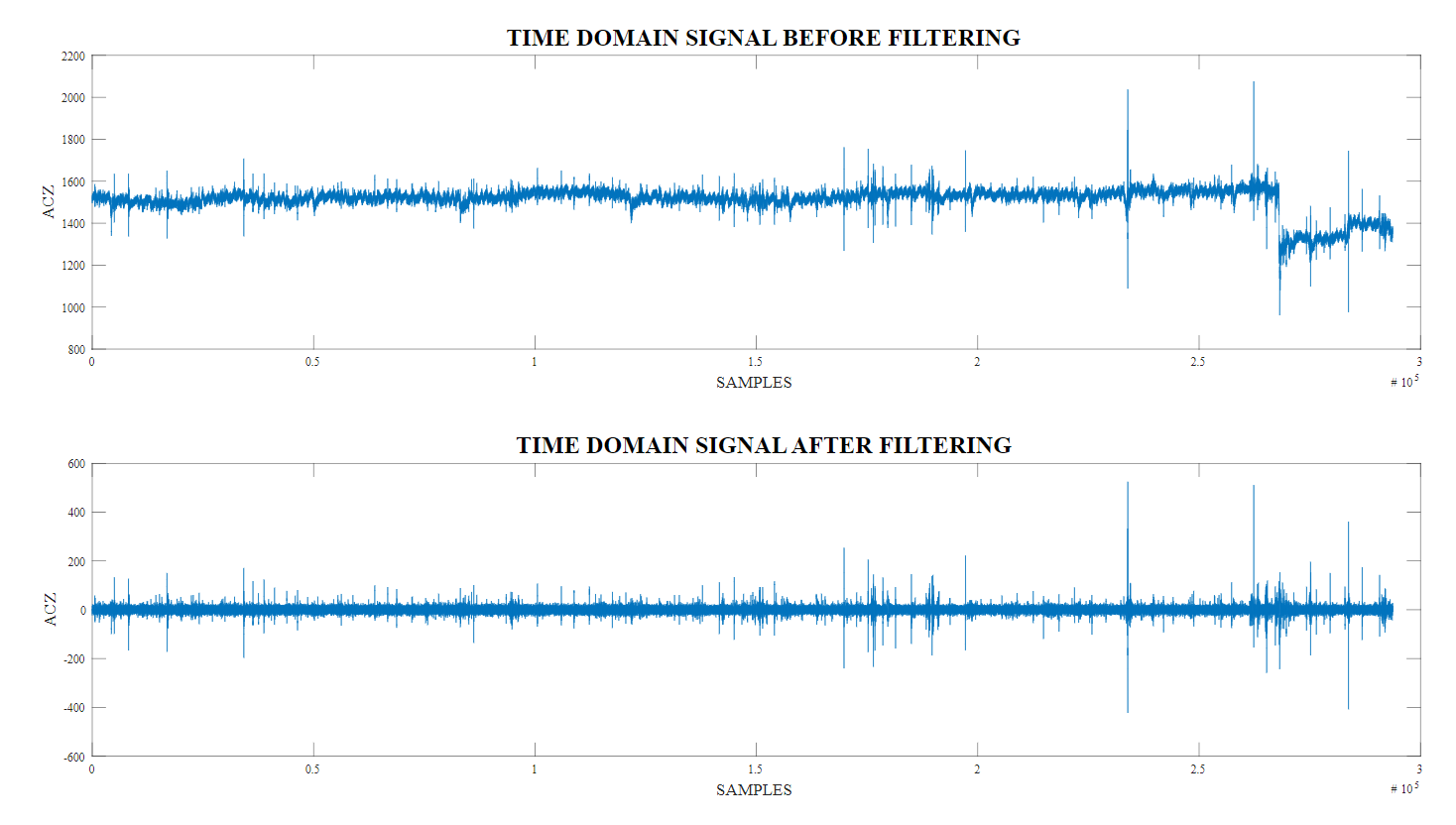}
\caption{{\bf The raw time domain data captured from the sensor Vs the time domain data when a custom high pass filter was applied}\
}
\label{fig:filtered}
\end{figure}

\begin{figure}[!h]
\includegraphics[width=\textwidth]{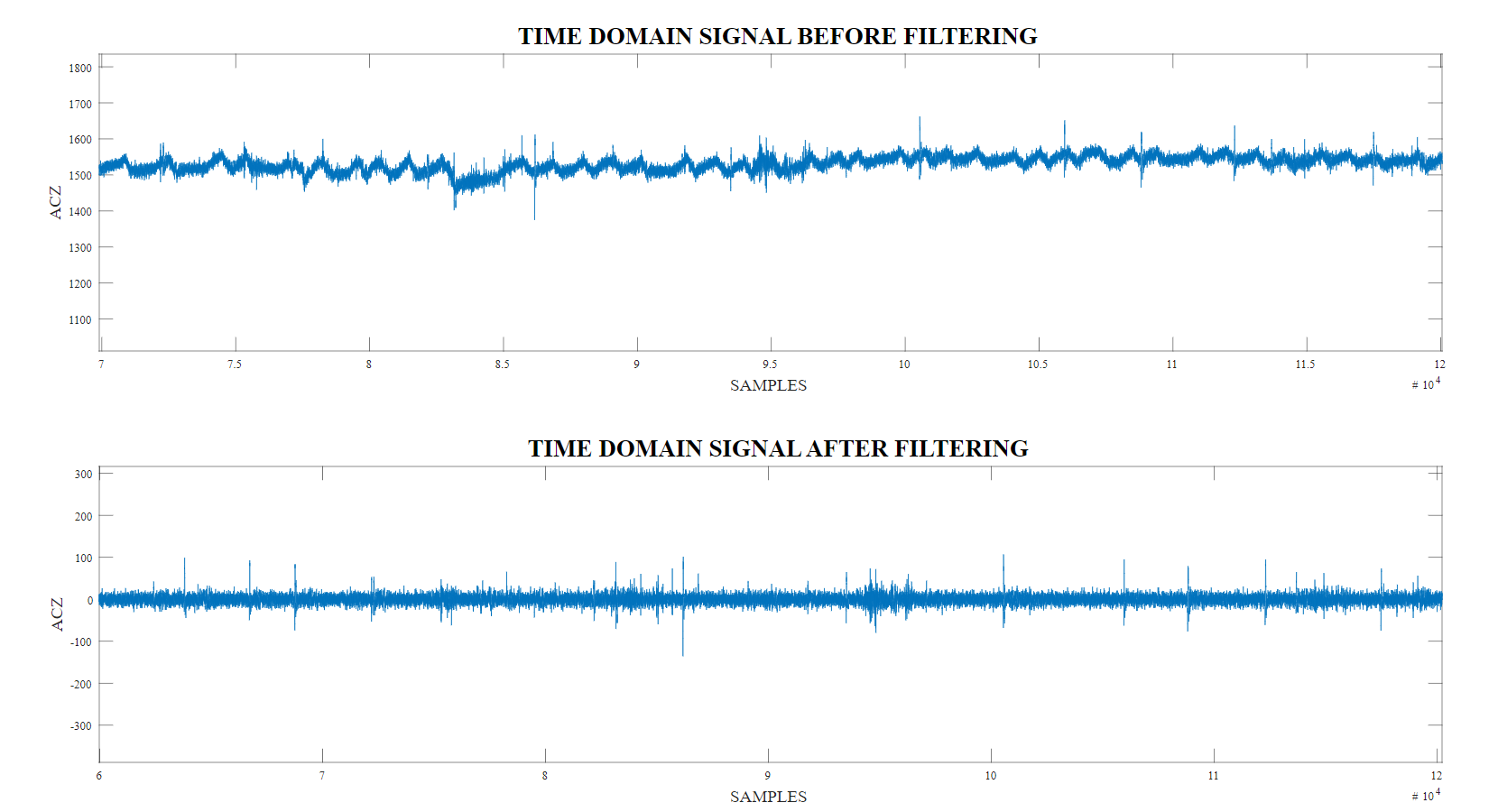}
\caption{{\bf The raw time domain data captured from the sensor Vs the time domain data when a custom high pass filter was applied (Zoomed-in)}\
}
\label{fig:filtered-zoom}
\end{figure}

\textbf{Realization Segmentation :} The output signal from the pre processor was then segmented in to realizations. These segments are non overlapping and had a width of 200 samples. This sample size was selected by observing the average length of fetal movements and maternal laughs. The realizations were classified into three classes: fetal movement, maternal laugh and maternal respiratory movements. The type of the realizations were determined by the input from the ultrasound scan, the pregnant mother as well as the data recorded via the button system.

\textbf{Short Time Fourier Transform :}The raw accelerometric data are segmented into three classes: fetal movements, maternal respiratory movements and maternal laugh. Each signal category contains unique features. However, conducting only time domain analysis will inhibit the ability to extract these features in order to classify them in to each class. Therefore, it was required to visualize them in a more descriptive manner. Furthermore, the input for the subsequent algorithm is required to be two dimensional. To achieve this purpose, Short Time Fourier Transform (STFT) was utilized to generate Magnitude Spectrogram from the time domain accelerometric readings \cite{stft}. The Magnitude Spectrogram is one of the most expressive signal representation methods, because it represents the intensity plot of frequencies of a signal which varies with time\cite{stft1}.  
\clearpage

\textbf{Non-Negative Matrix Factorization :}Non-Negative Matrix Factorization (NNMF) is widely used in signal processing to reduce the dimension \cite{21}. Applications of NNMF range from  simple text document clustering to advanced biological data mining \cite{18},\cite{20}. It is an advanced signal processing technique employed to identify and extract hidden features from raw signals. The input to the NNMF algorithm should be a non-negative matrix and the algorithm will generate two low rank non-negative matrices. This basic NNMF algorithm can be mathematically illustrated as follows \cite{19}.

\begin{equation}
V \approx WH 
\end{equation}
Where, \(V,W,H \geqslant 0\)\\

In this study of fetal movements identification application, the magnitude spectrogram was utilized as the input non-negative matrix V. Following the NNMF algorithm two matrices are generated. They are the Basis Matrix W and the Activation Coefficient Matrix (Abundance Matrix) H. These two matrices are generated such that the Basis Matrix contains all the features of the given magnitude spectrogram while the Activation Coefficient Matrix contains the proportional factors of bases. 

\textbf{Convoluted Neural Network :}

Convoluted Neural Network is one of the most common classifications methods utilized in biomedical signal processing \cite{cnn1}\cite{cnn2}. In order to classify the three classes the output from the above algorithm was fed in to a Convoluted Neural Network. 80\% of the data set was utilized for training and the selection process was random. The size of a realization fed into the CNN algorithm was 64x26 pixels and images were RGB. Following parameters of the CNN algorithm were varied and optimum values for each layer was obtained. The values of individual layers can be observed in Table 3. When training the network RelU Activation function was utilized and three fully connected layers were implemented for each class. Stochastic gradient descent method was used to train the network.

\begin{table}[!ht]
\centering
\caption{
{\bf Parameters of the three fully connected layers implemented in the Convoluted Neural Network.}}
\begin{tabular}{|l+l+l|l|l|}
\hline

$ \bf Parameter $ & \bf Layer 1  & \bf Layer 2  & \bf Layer 3  \\ \thickhline
$ \bf Neuron\hspace{0.2cm}Size $ & $5\times3$ & $5\times2$ & $5\times2$ \\ \hline
$ \bf Number\hspace{0.2cm}of\hspace{0.2cm}Neurons $ & 60 & 50 & 40 \\ \hline
$ \bf Learning\hspace{0.2cm}Rate $ &  0.0001 & 0.0001 & 0.0001\\ \hline
$ \bf Epochs $ &  80 & 150 & 300  \\ \hline
\end{tabular}

\begin{flushleft} In this table the parameters varied during implementing the convoluted neural network can be observed. The optimum parameters of each layer are mentioned.
\end{flushleft}
\label{CNN}
\end{table}

\subsection*{Mobile Application}
The final aim of this research study is to provide pregnant mothers with a wearable non-invasive device which can count and monitor fetal movement reliably. Therefore, initially a wearable deice was designed and fabricated. Then further studies were carried out to identify the best digital implementation method. During these, the digital literacy of the country was analysed. For this purpose, data released from the Department of Census and Statistics, Sri Lanka were utilized and following observations were made\cite{17}. The overall computer literacy of the female population during the year 2019 was 28.3\%. However, the average age group most pregnant mothers fall into is 20 to 35 years. The computer literacy of this age group vacillated between about 50\%. Compared to computer literacy, digital literacy of females have a higher value which is around 40\%. The digital literacy of females around the age of 20 to 35 vacillates between around 75\%. Therefore, it can be concluded that more pregnant mothers in Sri Lanka have better digital literacy than computer literacy. Hence, the smart phone was chosen as the digital implementation method. Moreover, it was derived from the survey conducted, that most mothers would prefer to use a mobile application along with the device.

Therefore, as the final system, the data capturing was done via the device in Fig 1, and stored in a micro SD card. After a session the mother can transfer the data included in the SD card to their smart phone. Initially, although an attempt was made to conduct an analysis within the smartphone, it was not successful as common smartphones used by pregnant mothers lacked the computational capacity to run the algorithm. Furthermore, the size of the data file of a single session is compact. Therefore, as a solution to these problems the data is then transferred to an online cloud and the analysis was conducted remotely. At the end of the analysis the number of kicks recorded within the session will  be sent back to the mother as well as to her obstetrician. When designing the mobile application the software Android Studios was utilized and when conducting the analysis remotely the Wamp Server Software was used. The mobile application was designed in a manner that is attractive as well as user friendly. The mobile application will store and record the fetal movement patterns as well. The user interface of the mobile application can be seen in Fig 9.  

\begin{figure}[!h]
\includegraphics[width=\textwidth]{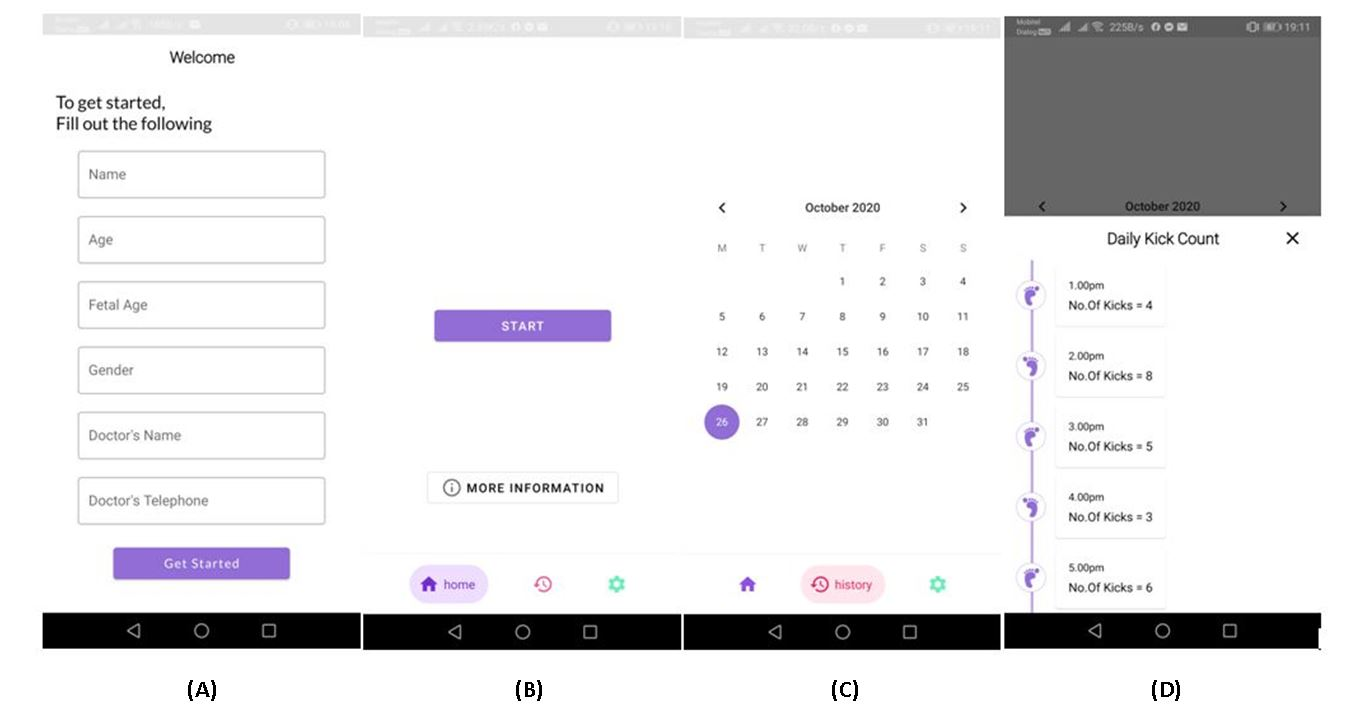}
\caption{{\bf Designed mobile application interface}\
}
\label{fig:app}
\end{figure}

\section*{Results}

During the clinical testing, readings from more than 120 mothers were taken. The distribution of gestational age and the fetal gender of the test group can be observed in Fig 10. From this figure, it can be observed that the gestation age of the test group varied from 27 weeks to 40+ weeks, where 40+ implies the fetus age is beyond 40 weeks old. Furthermore, it can be observed that the test group mostly contained fetuses whose gestation periods were beyond 37 weeks and the gender distribution was approximately uniform. However, the test group contained a higher number of younger male fetuses than female fetuses and a higher number of older female fetuses than male fetuses. There were few fetuses where the gender was not stated.

\begin{figure}[!h]
\includegraphics[width=\textwidth]{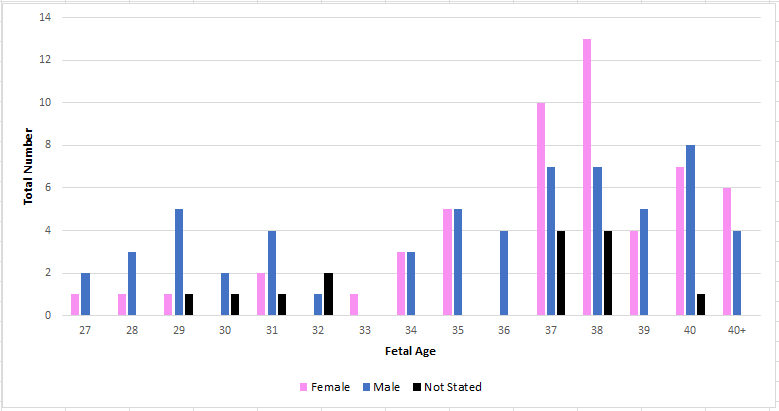}
\caption{{\bf Distribution of gestational age and the fetal gender of the test group}\
}
\label{fig:demography}
\end{figure}

The algorithms stated above were implemented on this data set and the results were observed. With the aim of obtaining an optimum algorithm, several combinations of algorithms were implemented. They are:

\begin{enumerate}
	\item{\textbf{Algorithm 1 :} Segmentation -- STFT -- CNN}
	\item{\textbf{Algorithm 2 :} High pass filter -- segmentation -- STFT -- CNN}
	\item{\textbf{Algorithm 3 :} High pass filter -- segmentation -- STFT --NNMF(W)-- CNN}
	\item{\textbf{Algorithm 4 :} High pass filter -- segmentation -- STFT --NNMF(H)-- CNN}
\end{enumerate}

In each algorithm immediately after the segmentation a short time Fourier transform was implemented. The resulting spectrograms can be seen in Fig 11. 
It can be observed that the spectrogram of the maternal respiratory movements is concentrated on to the lower bands of frequency, and this is constant through out the samples. However in the spectrogram of fetal movements this concentration occurs only in a smaller range of samples. Furthermore, the spectrogram of the maternal laugh signal differ from the other two as well. Therefore, it can be stated that implementing a standard short time Fourier transform can aid the discrimination process.

\begin{figure}[!h]
\includegraphics[width=\textwidth]{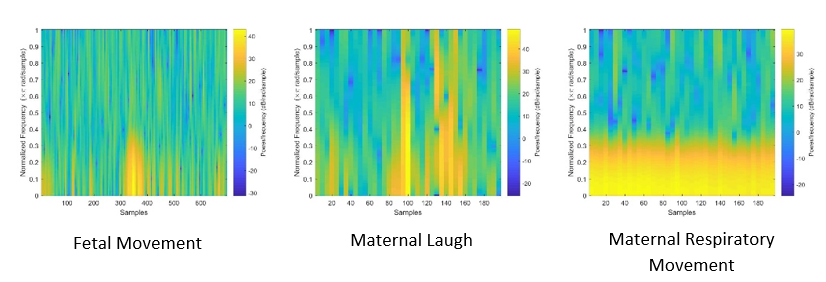}
\caption{{\bf The resulting spectrograms of the three classes}\
}
\label{fig:STFT}
\end{figure}

Then a standard Nonnegative Matrix Factorization algorithm was implemented on to the spectrogram images. This factorized the spectrogram image into two images: the abundance matrix and the base matrix. As stated in the previous section the abundance matrix is identified as the H matrix and the base matrix is identified as the W matrix for easy reference. Then the matrices are visualized in to figures with a grid format. Similar to the figures generated by the STFT algorithm, the amplitude of elements of the output matrices were converted in to colours and visualized. The generated figures for the W matrix of three different realizations can be observed in Fig 12 and the generated figures for the H matrix of the same three realizations can be observed in Fig 13.

\begin{figure}[!h]
\includegraphics[width=\textwidth]{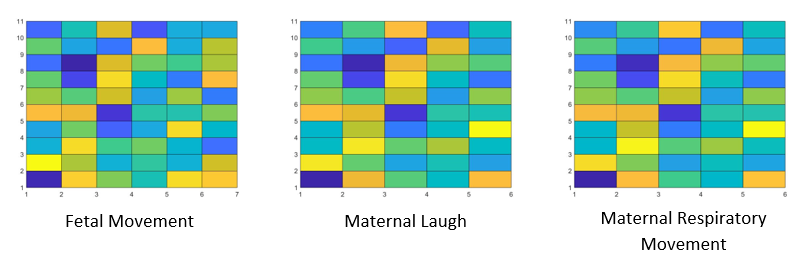}
\caption{{\bf The resulting base matrices (W matrices) for three classes}\
}
\label{fig:W}
\end{figure}
\begin{figure}[!h]
\includegraphics[width=\textwidth]{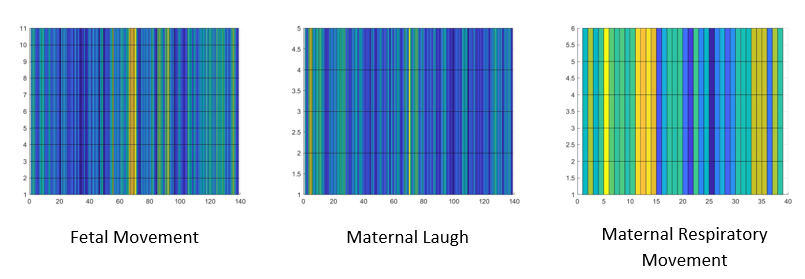}
\caption{{\bf The resulting abundance matrices (H matrices) for three classes}\
}
\label{fig:H}
\end{figure}

Then the spectrograms, and the base matrices and the abundance matrices were fed in to a standard convoluted neural network algorithm described above. From each following confusion matrices were obtained. In the given matrices Class 1 represents the fetal movement realizations, Class 2 represents the maternal laugh realizations and Class 3 represents the maternal respiratory movement realizations. And confusion matrices obtained by implementing the four algorithms are shown in Fig 14.

\begin{figure}[!h]
\includegraphics[width=\textwidth]{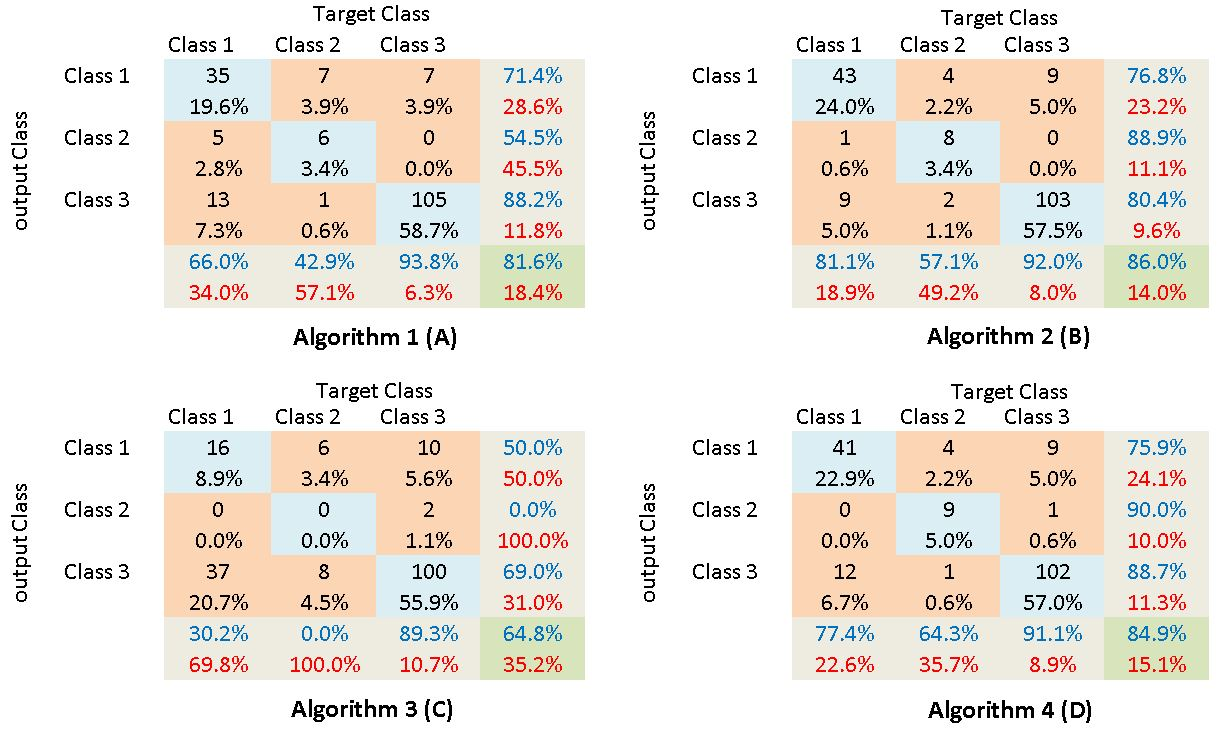}
\caption{{\bf The confusion matrices generated for each algorithm}\
}
\label{fig:conf}
\end{figure}

\section*{Discussion}

From the given confusion matrices following observations can be made. The confusion matrix in Fig 14 is obtained by implementing Algorithm 1, where the raw time-domain signal is fed into the STFT algorithm. From it, it can be observed while the accuracy of the algorithm is not at the best reasonable discrimination among classes can be obtained. The confusion matrix in Fig 15 is obtained by implementing Algorithm 2. In this initially, a high-pass filter was implemented to remove large maternal movements as well as maternal breathing pattern from the signal. Then the filtered signal was fed into the STFT algorithm and the remaining process is similar to Algorithm 1. By comparing the results obtained by implementing these two algorithms the effect of implementing a high-pass filter can be clearly observed. When comparing the two confusion matrices. it can be clearly observed that the true positive accuracy of Algorithm 2 is higher than the True positive accuracy of Algorithm 1.
However, the most important observation is the reduction of the false-positive rate with the implementation of the high pass filter. In this application of fetal movement monitoring, higher attention is paid to the false positive rate. When a false positive occurs the observer will identify a fetal movement when actually a fetal movement has not occurred. Therefore, if the false positive rate is higher then the application will indicate a higher fetal movement rate than the actual. Therefore, mothers and health care providers won't be able to identify a significant reduction in fetal movement rates early, which may lead to dire consequences. It can be observed that in Algorithm 2 the false positive rate is comparatively low than the false positive rate of Algorithm 1. From these observations, it can be observed that the introduction of the high-pass filter as a pre-processing step has improved the results of the algorithm.

The confusion matrices obtained by implementing Algorithm 3 and Algorithm 4 can be observed in Fig 16 and Fig 17 respectively. In Algorithm 2 the entire spectrogram was fed into the CNN algorithm. However, in Algorithm 3 and 4, the spectrogram was factorized using the Non-Negative Matrix factorization and the resulting abundance matrices and the base matrices are fed in the CNN respectively. The main reason to factorize the spectrogram is to reduce the computational capacity required to run the CNN algorithm. However, the following observations were made by analyzing the confusion matrices. When considering the observations made when Algorithm 3 is implemented it can be noted that the performance of this algorithm is weak. The true positive rate is at the lowest of 65\% and the false positive rate is at it's highest of 9\%.

However, the best results were obtained when the spectrograms are fed into the CNN rather than any factorized matrices. The true positive rate of Algorithm 2 is 86\% and the false positive rate is around 7\%. The next best result is observed when the (Abundance Matrix) H is fed into the CNN algorithm. The true positive rate of it is approximately 85\% and the false positive rate is around 7.2\%. When comparing these two algorithms the decrease in the true positive rate can be neglected. However, the false-positive rate will increase due to factorization. For the same explanation made above, in this application, more attention needs to be paid to the effect of the false-positive rate.

The main reason for the better performance of Algorithm 2 could be due to the properties of the factorization algorithm. When the spectrogram matrix is factorized into the abundance matrix and the base matrix the abundance matrix contains mainly the data pertaining to individual mothers. Therefore, when the abundance matrices are fed into the CNN algorithm since the algorithm is trained utilizing data obtained from several mothers the performance is low. However, if we were to train the network using the abundance matrices of an individual mother the performance may be better. This can be observed in Table 4 where Algorithm 2 to 4 are implemented on 6 individual mothers and the true positive rates are compared. In the table, it can be observed that for some mothers the true positive rate obtained using Algorithm 4 is higher. This solidifies the argument made above. However, training a network utilizing an individual mother is not feasible for the intended application. Furthermore, if the training is to be done to individual mothers a large number of samples need to be collected from individual mothers. This does not go along with the target of devising a universal system to monitor fetal movements. Furthermore, it can be observed the performance of Algorithm 3 is weak even if it was implemented on individual mothers.

\begin{table}[!ht]
\centering
\caption{
{\bf True Positive rate of Algorithm 2 to 4 on individual mothers..}}
\begin{tabular}{|l+l|l|l|l|}
\hline
$ \bf Mother Index $ & \bf A2 (\%) & \bf A4(\%) & \bf A3(\%) \\ \thickhline
$ \bf 1 $ & 93.55 & 83.87 &  41.94\\ \hline
$ \bf 2 $ & 88.89 & 88.89 & 63.49\\ \hline
$ \bf 3 $ & 78.57 & 71.43 & 64.24\\ \hline
$ \bf 4 $ & 71.43 & 71.43 &  42.86\\ \hline
$ \bf 5 $ & 71.43 & 71.43  & 57.14\\ \hline
$ \bf 6 $ & 68.75 & 75.00 & 43.75\\ \hline
\end{tabular}
\begin{flushleft} In this table the true positive rates observed when the three algorithms are implemented on six individual mothers. 'A2' refers to Algorithm 2 , 'A4' refers to Algorithm 4 and 'A3' refers to Algorithm 3. The results of Algorithm 2 and Algorithm 4 are juxtaposed for the purpose of better discrimination among the two. The results from Algorithm 3 are also included.
\end{flushleft}
\label{TPR}
\end{table}

When all the observations mentioned above are considered it can be concluded that the best algorithm in Algorithm 2, where initially a high pass filter is implemented, then a spectrogram was obtained and it was fed into a Convoluted Neural Network Algorithm. Furthermore, all the computations are being carried out in an online cloud. This provides a higher computational capacity, which helps when the spectrogram is directly fed into the CNN algorithm.

\section*{Conclusion}
The need for a proper system to monitor fetal movements in a non-clinical setup is of paramount importance in order to maintain fetal well-being. Although some previous research studies were conducted on this front, several crucial and novel concepts were introduced in this study. A complete system was introduced to be used by pregnant mothers to monitor fetal movement count in a non-clinical setting. While a significant amount of effort was spent on developing the algorithm as well as the sensing system an equal amount of effort was invested in designing and implementing proper ergonomics and a user-friendly interface to the system. It was made sure that the proposed system is feasible to be implemented. This was done by studying the preferences and habits of pregnant mothers. Furthermore, it was ensured that the system is user friendly in nature. This was aided by the extensive amount of surveys conducted while clinical testing procedures.

In the initial phases of the research, a proper non-invasive device was designed and fabricated. Then from the feedback received from pregnant mothers, it was further modified. Subsequently, a mobile application was developed to be used by mothers. Finally, four different algorithms were implemented on the data set to identify the most acceptable algorithm. Algorithm 2 gave out the most promising results while the performance of Algorithm 4 was also close to Algorithm 2. Furthermore, due to the facts such as data file size and lack of computational capacity limitations Algorithm 2 was selected to be the most befitting algorithm to the system.

In conclusion, in this research, a low-cost, non-transmitting wearable system was designed to monitor fetal movement patterns. This system consists of a non-invasive sensing unit, a mobile application to be used by pregnant mothers, and an algorithm to extract the required information from the captured data. This system would be of immense use for pregnant mothers as well as for researches who would be required to collect data to analyze fetal movement patterns.

\section*{Acknowledgments}

We would like to express our gratitude to all the participants of the clinical testings for volunteering as well as for providing feedback on the proposed system. We also would like to thank the healthcare professional at the Peradeniya Teaching Hospital for all the help and guidance provided. We also express our gratitude to our reviewers for their constructive feedback.

%
%
%

\end{document}